\documentclass[runningheads]{svmult}

\usepackage{makeidx}   % allows index generation
\usepackage{graphicx}  % standard LaTeX graphics tool
\usepackage{subeqnar}  % subnumbers individual equations
\usepackage{multicol}  % used for the two-column index
\usepackage{physprbb}  % modified textarea for proceedings,
\makeindex             % used for the subject index
\begin{document}
\title*{Galaxy Clusters: Cosmic High-Energy Laboratories to 
Study the Structure of Our Universe}
\toctitle{Galaxy Clusters: Cosmic High-Energy Laboratories}
% allows explicit linebreak for the table of content
%
%
\titlerunning{Galaxy Clusters: Cosmic High-Energy Laboratories}
% allows abbreviation of title, if the full title is too long
% to fit in the running head
%
\author{Hans B\"ohringer}
\authorrunning{Hans B\"ohringer et al.}
% if there are more than two authors,
% please abbreviate author list for running head
%
%
\institute{Max-Planck-Institut f\"ur Extraterrestrische Physik,         
D-85748 Garching, Germany}

\maketitle              % typesets the title of the contribution

\begin{abstract}
This contribution illustrates the study of galaxy clusters as
astrophysical laboratories as well as probes for the large-scale
structure of the Universe. Using the REFLEX Cluster Survey, the
measurement of the statistics of the large-scale structure on scales up
to 500 $h^{-1}$ Mpc is illustrated. The results clearly favour a low 
density Universe.

Clusters constitute, in addition, well defined astrophysical laboratory
environments in which some very interesting large-scale phenomena can
be studied. As an illustration we show some spectacular new XMM
X-ray spectroscopic results on the thermal structure of cooling
flows and the interaction effects of AGN with this hot intracluster medium.
The X-ray observations with XMM-Newton show a lack of spectral evidence
for large amounts of cooling and condensing gas in the centers of 
galaxy clusters believed to harbour strong cooling flows.
To explain these findings we consider
the heating of the core regions of clusters by jets from a central 
AGN. We find that the power output the AGN jets is well sufficient.
The requirements such a heating model has to fulfill are explored
and we find a very promising scenario of self-regulated Bondi accretion
of the central black hole.
\end{abstract}

\section{Introduction}
Galaxy clusters with masses from about $10^{14}$ to over $10^{15}$
M$_{\odot}$ are the largest clearly defined building blocks
of our universe. Their formation and evolution is tightly connected
to the evolution of the large-scale structure of our Universe as a
whole. Clusters are therefore ideal probes for the study of
the large-scale matter distribution.
Due to the hot, intracluster plasma trapped in the potential well
of galaxy clusters, which can take on temperatures of several keV
(several 10 Million degrees), galaxy clusters are the most luminous
X-ray emitters in the Universe next only to quasars. The hot gas
and its X-ray emission is a good tracer of the gravitational
potential of the clusters and thus allows us to obtain mass estimates
of the clusters, to study their morphology and dynamical state,
and to detect clusters as gravitationally bound entities out to
very large distances (Sarazin 1986). Systematic studies have shown
that clusters have within a first order description, a quite 
standardized, self-similar appearance. This is reflected
observationally in quite narrow correlations of observable parameters,
like e.g. the temperature - X-ray luminosity relation. The for our
application most important relation is that of mass and X-ray
luminosity. Since, for the construction of a cosmologically
interesting cluster sample, we are interested to collect all the
clusters above a certain mass limit, this relation makes it possible
to select the most massive clusters essentially by their X-ray luminosity.   

\begin{figure}
\begin{center}
\includegraphics[width=0.70\textwidth]{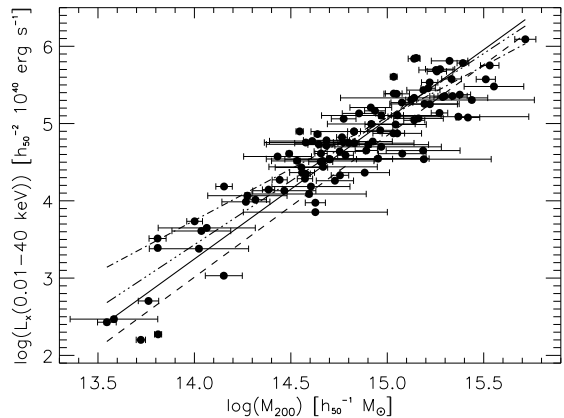}
\end{center}
\caption{Mass-X-ray luminosity relation determined from
the 106 brightest galaxy clusters found in
the ROSAT All-Sky Survey by Reiprich \& B\"ohringer (1999, 2001)}
\label{Mfunc}
\end{figure}

The first detailed mass-X-ray luminosity relation compiled from X-ray
observations is shown in Fig.1 (Reiprich \& B\"ohringer 1999, 2001).
From these results a cluster mass function and its integral, the
matter density in the Universe bound in clusters was
derived. In Fig. 2 we show this cumulative cluster mass density function, yielding
$\Omega_{clusters}\sim 0.02$. Thus for the currently most popular
value for the matter density $\Omega_m \sim 0.3$ about 6\% of the
matter in the Universe is bound in clusters with a mass larger than
$6.4 \cdot 10^{13} h_{50}^{-1}$ M$_{\odot}$.

\begin{figure}
\begin{center}
\includegraphics[width=0.70\textwidth]{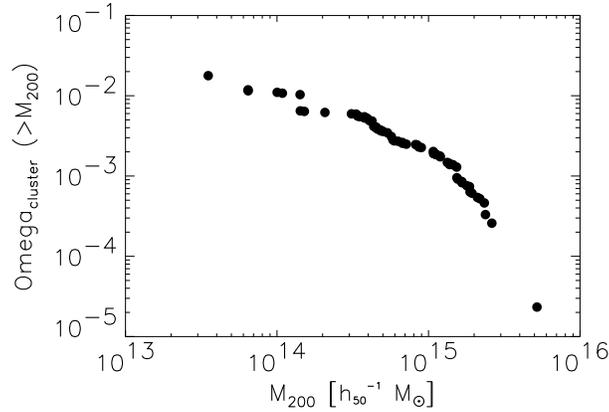}
\end{center}
\caption{Cumulative fraction of the mass density bound in
clusters in the Universe compared to the critical density (Reiprich \& 
B\"ohringer 2001)}
\label{Mfunc}
\end{figure}

\section{Measurement of the Large-Scale Structure}
\label{sec:surveys}
Based on
the ROSAT All-Sky X-ray Survey - so far the only all-sky survey conducted
with an X-ray telescope - we exploit the above mentioned tight
X-ray luminosity-mass relation with X-ray
selected cluster redshift surveys to study the large-scale structure
of the Universe. Fig.3 shows the sky
distribution of the brightest
X-ray galaxy clusters identified so far in the ROSAT All-Sky Survey
within two projects: the northern NORAS Survey
(B\"ohringer et al. 2000) and the southern REFLEX
Survey (B\"ohringer et al. 2001a). The latter project is currently
more complete comprising a sample of 452 clusters and therefore
the following results are based
on this project. Most of the redshifts for this survey have
been obtained within the frame of an ESO key program
(B\"ohringer et al. 1998, Guzzo et al. 1999) .

\begin{figure}
\begin{center}
\includegraphics[width=0.65\textwidth]{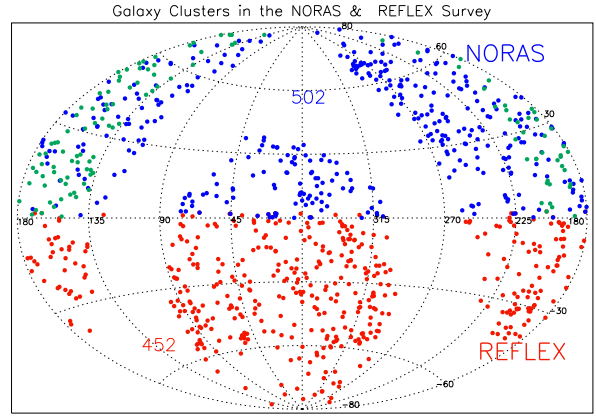}
\end{center}
\caption{
Sky distribution of the brightest galaxy clusters found in the
ROSAT All-Sky Survey investigated in the NORAS and REFLEX redshift surveys.}
\label{fig:skydistribution}
\end{figure}

The basic census of the REFLEX cluster survey is the X-ray luminosity
function shown in Fig. 4 (B\"ohringer et al. 2002).
The essential goal of this survey is the assesment of the
statistics of the large-scale structure.
The most fundamental statistical description of the spatial structure
is based on the second moments on the distribution,
characterized either by the two-point-correlation function or its Fourier
transform, the density fluctuation power spectrum. The two-point
correlation function of the clusters in
the REFLEX sample has been derived by Collins et al. (2000). 
The results show
a power law shaped correlation function with a slope of $ 1.83$, 
a correlation length of
$18.8 h_{100}^{-1}$ Mpc and a possible zero crossing at
$ \sim 45 h_{100}^{-1}$ Mpc. The density fluctuation power spectrum
(Fig. 5) has been determined by Schuecker et al. (2001a).
The power spectrum is characterized by a power law at
large values of the wave vector,
$k$, with a slope of $\propto k^{-2}$ for $k \le 0.1 h$ Mpc$^{-1}$
and a maximum around $k \sim 0.03 h$ Mpc$^{-1} $
(corresponding to a wavelength of
about $200 h^{-1}$ Mpc). This maximum reflects
the size of the horizon when the Universe featured equal
energy density in radiation
and matter and is a sensitive measure of the
mean density of the Universe, $\Omega_0$.
For standard open and flat Cold Dark Matter models
(OCDM and $\Lambda$CDM) we find the following constraints
$h\Omega_0 = 0.12~ {\rm to}~ 0.26$ (Schuecker et al. 2001a).

\begin{figure}
\begin{center}
\includegraphics[clip=,width=0.67\textwidth]{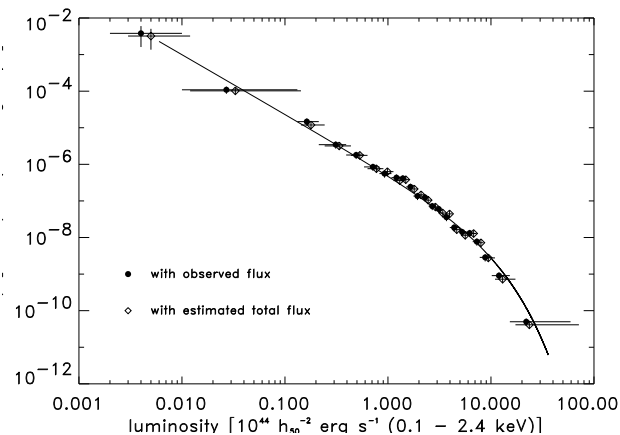}
\end{center}
\caption{
X-ray luminosity function of the REFLEX cluster Survey (B\"ohringer et
al. 2002). The two series 
of data point refer to the calculation of the luminosities
from the X-ray fluxes as observed and corrected for the missing flux
in the outer very low surface brightness regions, respectively.}

\label{fig:rxlf}
\end{figure}

\begin{figure}
\begin{center}
\includegraphics[clip=,width=0.67\textwidth]{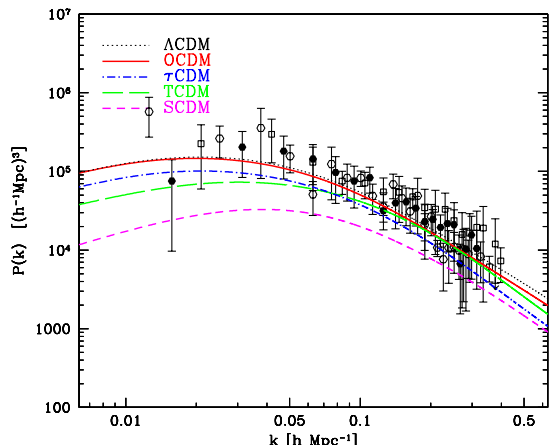}
\end{center}
\caption{
Power spectra of
the density fluctuations in the REFLEX cluster
sample together with predictions from various popular cosmological models
taken from the literature. For details see Schuecker et al. (2001a).
}
\label{fig:rpk}
\end{figure}

\section{Cosmological Tests with Galaxy Clusters}

In addition to the above mentioned large-scale structure test for
cosmological models several further studies yield information on
cosmological parameters. These include the modelling of  
the mass, temperature, and luminosity
functions (e.g. Reiprich \& B\"ohringer 2001, Ikebe et al. 2001), the
study of the mass-to-light ratio in connection with galaxy biasing
Carlberg et al. (1996; which yields $\Omega_0 \sim 0.24 \pm 0.1$),
the baryon mass fraction in clusters,
and the statistics of merging clusters as
observed in the REFLEX cluster sample (Schuecker et al. 2001b) .
All these different cosmological studies using clusters point towards a low density
universe. 
These results can be compared to the results obtained from
observations of the cosmic microwave anisotropies
(e.g. De Bernardis et al. 2000, 2001)  
and of the study of
distant SN Ia (e.g. Perlmutter et al. 1999). While these two investigations
provide combined constraints that encircle a region in the model parameter
space spanned by the cosmological parameters $\Omega_0$ and
$\Omega_{\Lambda}$ around values of  $\Omega_0 = 0.3$ and
$\Omega_{\Lambda} = 0.7$, the galaxy cluster results provide a different
cut through this parameter space crossing the other two results at
their intersection. That is, the cluster results provide at present
(without the inclusion of an investigation of a very large redshift range)
no significant constraints on the $\Omega_{\Lambda}$-parameter, but 
allow values of for $\Omega_0 = 0.2 - 0.4$ in the range consistent
with the combined cosmic microwave and SN Ia data. Thus,
the evidence for a low density universe is solidifying.

\section{Cluster Cooling Cores}
X-ray imaging observations have shown
that the X-ray emitting, hot gas in a large fraction of all galaxy clusters
reaches high enough densities in the cluster centers that the cooling
time of the gas falls below the Hubble time,
and gas may cool and condense
in the absence of a suitable fine-tuned heating source
(e.g. Silk 1976, Fabian \& Nulsen 1977).
From the detailed analysis of surface brightness profiles of X-ray images of
clusters obtained with the {\sl Einstein}, {\sl EXOSAT},
and {\sl ROSAT} observatories,
the detailed, self-consistent scenario of inhomogeneous, comoving
cooling flows emerged (e.g. Nulsen 1986, 
Thomas, Fabian, \& Nulsen 1987, Fabian 1994). The main 
assumptions on which the cooling flow model is based and some important 
implications are: (i) Each radial zone
in the cooling flow region comprises different plasma phases
covering a wide range of temperatures. The consequence
of this temperature distribution is that gas will cool to low 
temperature and condense over a wide range of radii. 
(ii) The gas features an inflow in
which all phases with different temperature
move with the same flow speed.
(iii) There is no energy exchange between the different phases,
between material at different radii, and no heating. 

Now the first analysis of high resolution
X-ray spectra and imaging spectroscopy obtained with {\sl XMM-Newton} 
has shown to our surprise
that the spectra show no signatures of cooler phases of the 
cooling flow gas below an intermediate temperature 
(e.g. Peterson et al. 2001, Tamura et al. 2001) and 
local isothermality in the cooling flow region 
(e.g. B\"ohringer et al. 2001b, Matsushita
et al. 2001, Molendi \& Pizzolato 2001) in conflict with the
inhomogeneous cooling flow model. 
Here, we discuss these new spectroscopic results and their
implications and point out the way to a new possible
model for this phenomenon. The results are mostly based on the 
detailed observations
of the M87 X-ray halo. A detailed description of this study is
provided by B\"ohringer et al. (2001c).

\section{Spectroscopic Diagnostics of Cluster Cooling Cores}

XMM Reflection Grating Spectrometer (RGS) observations of
several cooling core regions show
signatures of different temperature
phases ranging approximately from the hot virial temperature of
the cluster to a lower limiting temperature, $T_{low}$.
Clearly observable spectroscopic features of even
lower temperature gas expected for a cooling flow model are
not observed. A1835 with a bulk temperature of about
8.3 keV has  $T_{low}$ around 2.7 keV (Peterson et al. 2001) and
similar results have been derived for A1795 (Tamura et al. 2001).
These results are very well confirmed by {\sl XMM} observations
with the energy sensitive imaging devices, EPN and EMOS,
providing spectral
information across the entire cooling core region,
yielding the result that (for
M87, A1795, and A1835)
single temperature models provide a better representation
of the data than cooling flow models (B\"ohringer et al. 2001a,
Molendi \& Pizzolato 2001) also implying the lack of low temperature
components. The very detailed analysis of M87 by
Matsushita et al. (2001 and the contribution to this workshop) 
has shown that the temperature
structure is well described locally by a single temperature
over most of the cooling core region, except for 
the regions of the radio lobes
and the very center ($r \le 1$ arcmin, $\sim 5$ kpc).

\begin{figure}[h]
\begin{center}
\includegraphics[width=0.9\textwidth]{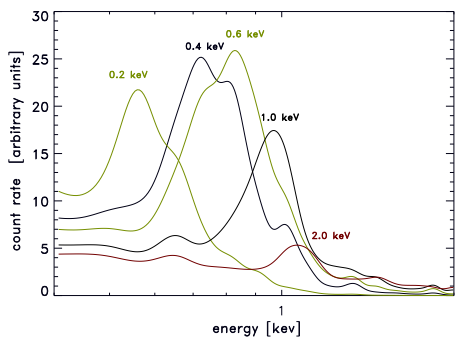}
\end{center}
\caption[]{The Fe L-line complex in X-ray spectra as a function of the plasma
temperature for a metallicity value of 0.7 solar. The simulations
show the appearance of the spectra as seen with the XMM EPN.
The emission measure was kept fixed when the temperature was varied.}
\label{eps1}
\end{figure}

Among the spectroscopic signatures which are sensitive to the 
plasma temperature in the relevant temperature range,
the complex of iron  L-shell lines is most
important. 
Fig.\ 6 shows simulated X-ray spectra 
as predicted for the XMM EPN instrument in the spectral
region around the Fe L-shell lines for a single-temperature plasma
at various temperatures from 0.4 to 2.0 keV and 0.7 solar metallicity.
There is a very obvious shift in the location
of the peak making this feature an excellent
thermometer.
For a cooling flow with a broad range of temperatures one expects a composite
of several of the relatively narrow line blend features, resulting in 
a quite broad peak.
Fig.\ 7a shows for example the deprojected 
spectrum of the M87 halo plasma for the radial range 1 - 2 arcmin
(outside the inner radio lobes) and
a fit of a cooling flow model with a mass deposition rate slightly less
than 1 M$_{\odot}$ yr$^{-1}$ as expected for this radial
range from the analysis of the surface
brightness profile (e.g. 
Matsushita et al. 2001). 
It is evident that the peak
in the cooling flow model is much broader than the observed spectral
feature. For comparison Fig.\ 7b shows the same spectrum fitted
by a cooling flow model where a temperature of 2 keV was chosen for
the maximum temperature and a suitable lower temperature cut-off
(1.44 keV) was
determined by the fit. The very narrow temperature interval (almost
isothermality) is well consistent with the narrow peak. A similar
result is obtained for other clusters, e.g. A1795
as shown in Fig.\ 7d.

\begin{figure}[b]
\begin{center}
\includegraphics[width=1.0\textwidth]{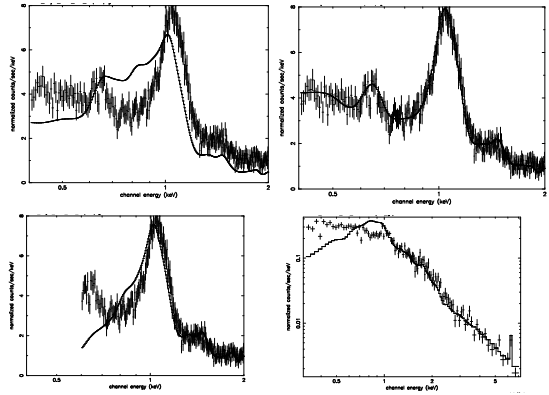}
\end{center}
\caption[]{XMM EPN spectrum 
of the M87 X-ray halo in the radial range $R=1 - 2$ arcmin. The
spectrum has been fitted with a cooling flow model with a best fitting
mass deposition rate of 0.96 M$_{\odot}$ yr$^{-1}$ and a fixed
absorption column density of
$1.8 \cdot 10^{20}$ cm$^{-2}$, the galactic value, and a parameter
for $T_{low}$ of 0.01 keV.
{\bf (b - upper right):} same spectrum fitted by a cooling flow spectrum
artificially constraint to emission from the narrow 
temperature interval 1.44 - 2.0 keV, where $T_{low}$ was treated as
a free fitting parameter.
{\bf (c - lower left):} same spectrum fitted with a free parameter
for the internal excess absorption. The spectrum was constraint to
the energy interval 0.6 to 2.0 keV.
{\bf (d - lower right):} {\sl XMM} EPN spectrum of A1795 fitted with a
cooling flow model with the galactic value for absorption.}
\label{eps2}
\end{figure}

Since this diagnostics of the temperature structure is essentially
based on the observation of metal lines, an inhomogeneous
distribution of the metal abundances in the cluster ICM
and a resulting suppression of line emission at low temperatures
was suggested as a possible way to reconcile
the above findings with the standard cooling flow model
by Fabian et al. (2001 and contribution in these proceedings).
As shown by B\"ohringer et al. (2001c) such a scenario  will still
result in a relatively broad Fe L-line feature
and does not solve the problem in this case of M87.

\section{Internal absorption}

Another possible attempt to obtain consistency
is to allow the absorption 
parameter in the fit to adjust freely. This is demonstrated in 
Fig.\ 7c with the same observed spectrum where
the best fitting absorption column
density is selected in such a way by the fit that the absorption edge
limits the extent of the Fe-L line feature towards lower energies.
This is actually the general finding with ASCA observations
showing two possible options for the
interpretation of the spectra of cluster core regions: (1) an interpretation
of the results in form of an inhomogeneous cooling flow model which
than necessarily includes an internal absorption component
(e.g. Allen 2000, Allen et al. 2001), or (2)
an explanation of the spectra in terms of a two-temperature component
model (e.g. Ikebe et al. 1999, Makishima 2001) where the hot 
component is roughly equivalent to the hot bulk temperature of the
clusters and the cool component corresponds approximately to
$T_{low}$. Thus for the cooling flow interpretation to work and
to produce a sharp Fe-L line feature as observed, the absorption edge
has to appear at the right energy and therefore values for the 
absorption column of typically around $3 \cdot 10^{21}$ cm$^{-2}$
are needed (e.g. Allen 2000 and Allen et al. 2001 who find
values in the range $1.5 - 5 \cdot 10^{21}$ cm$^{-2}$). 

It is therefore important to perform an independent test on the presence of
absorbing material in the cluster cores.
Thanks to {\sl CHANDRA} and {\sl XMM-Newton} we can now use central
cluster AGN as independent light sources for probing.
Using the nucleus and jet of M87 (with {\sl XMM}, see Fig. 3)
and the nucleus of
NGC1275 (with {\sl CHANDRA}) we find no signature of internal
absorption. Thus at least for these two cases 
internal absorption is not observed.

\begin{figure}[b]
\begin{center}
\includegraphics[width=.70\textwidth]{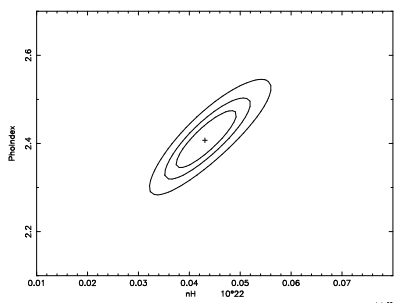}
\end{center}
\caption[]{Constraints on the shape of the {\sl XMM EPN}-spectrum of 
the nucleus of M87. The lines show the 1, 2, and 3$\sigma$ confidence
intervals for the combined fit of the slope (photon index) of the
power law spectrum and the value for the absorbing column density,
$n_H$, in units of $10^{22}$ cm$^{-2}$.}
\label{eps3} 
\end{figure}

\section{Heating Model}

In view of these difficulties of interpreting the observations
with the standard cooling flow model, we may consider
the possibility that the cooling and mass deposition rates
are much smaller than previously thought, that is reduced by
at least one order of magnitude.
To decrease the mass condensation under energy
conservation some form of heating is clearly necessary.
Three forms of heat input into the cooling flow
region have been discussed: (i) heating by the energy output of the
central AGN (e.g.
Pedlar et al. 1990, Tabor \& Binney 1993, McNamara et al. 2000,
(ii) heating by heat conduction from the hotter gas 
outside the cooling flow (e.g. Tucker \& Rosner 1983, Bertschinger \&
Meiksin 1986), and (iii) heating by magnetic fields, basically through some 
form of reconnection (e.g. Soker \& Sarazin 1990, Makishima et al. 2001).
The latter two processes depend on poorly known plasma physical
conditions and are thus more speculative.
The energy output of the central AGN, however, can be determined
as shown below.

A heating scenario can only successfully explain the observations
if among others the two most important requirements are met:
(i) The energy input has to provide sufficient heating
to balance the cooling flow losses, that is about $10^{60}$ to 
$10^{61}$ erg in 10 Gyr or on average about $3 \cdot 10^{43} -
3\cdot 10^{44}$ erg s$^{-1}$, and (ii)
The energy input has to be fine-tuned. Too much heating 
would result in an outflow from the central region and the central
regions would be less dense than observed. Too little heat will
not reduce the cooling flow by a large factor. Therefore the 
heating process has to be self-regulated: mass deposition triggers
the heating process and the heating process reduces the mass 
deposition.

Further constraints are discussed by B\"ohringer et al. (2001c).
The total energy input into the ICM
by the relativistic jets of the central AGN can be
estimated by the interaction effect of the jets with
the ICM by means of the scenario described in Churazov et al. (2000).
It relies on a comparison of the inflation and 
buoyant rise time of the bubbles of relativistic plasma
which are observed e.g. in the case of NGC 1275 (B\"ohringer et al. 1993,
Fabian et al. 2000). The estimated total energy output is
for three examples, M87: $1.2 \cdot 10^{44}$ erg s$^{-1}$, 
Perseus: $1 \cdot 10^{45}$ erg s$^{-1}$, 
and Hydra A: $2 \cdot 10^{45}$ erg s$^{-1}$.
These values
for the energy input have to be compared with the energy loss in the 
cooling flow, which is of the order of $10^{43}$ erg s$^{-1}$
for M87 and about  $10^{44}$ erg s$^{-1}$ for Perseus.
Thus in these cases the energy input
is larger than the radiation losses in the cooling flow 
for at least about the last $10^8$ yr.
We have, however, evidence that this energy input continued
for a longer time with evidence given by the outer radio halo
around M87 with an outer radius of 35 - 40 kpc (e.g. Kassim et al.
1993, Rottmann et al. 1996). Owen et al. (2000) give a
detailed physical account of the halo and model the energy input
into it. They estimate the total current energy content in the halo
in form of relativistic plasma to $3 \cdot 10^{59}$ erg and
the power input for a lifetime of about $10^8$ years, which is also
close to the lifetime of the synchrotron emitting electrons,
to the order of $10^{44}$ erg s$^{-1}$, consistent with our estimate.
The very characteristic sharp outer boundary of the 
outer radio halo of M87,
noted by Owen et al. (2000), has the important
implications, that this could not have been produced by magnetic
field advection in a cooling flow.

Thus, we find a radio structure providing evidence for a
power input from the central AGN into the halo region 
of the order of about ten times
the radiative energy loss rate over at least about $10^8$ years
(for M87).
The energy input could therefore balance the heating for at least
about $10^9$ years.
The observation of active AGN in the centers of cooling
flows is a very common phenomenon. E.g. Ball et al. (1993)
find in a systematic VLA study of the radio properties of cD galaxies
in cluster centers, that 71\% of the cooling flow clusters have
radio loud cDs compared to 23\% of the non-cooling flow cluster cDs.
Therefore we can safely assume that the current
episode of activity was not the only one in the life of M87 and its
cooling flow.

The mechanism for a fine-tuned heating of the cooling flow region
should most probably be searched for in a feeding mechanism of the
AGN by the cooling flow gas. The most simple physical situation
would be given if simple Bondi
type of accretion from the inner cooling core region would roughly
provide the order of magnitude of the power output that is observed
and required.
Using the classical formula for spherical accretion from a hot gas
by Bondi (1952) we can obtain a very rough estimate for this number.
For the proton density near the M87 nucleus ($r \le 15$ arcsec)
of about 0.1 cm$^{-3}$, a temperature of about $10^7$ K
(e.g. Matsushita et al. 2001), and
a black hole mass of $3\cdot 10^9$ M$_{\odot}$ (e.g. Ford et al.
1994) we find a
mass accretion rate of about 0.01
M$_{\odot}$ yr$^{-1}$ and an energy output of about $7 \cdot 10^{43}$
erg s$^{-1}$, where we have assumed the canonical value of 0.1 for the ratio
of the rest mass accretion rate to the energy output.
The corresponding accretion radius is about 50 pc ($\sim 0.6$ arcsec).
This accretion rate is more than a factor of 1000 below the Eddington
value and thus no reduction effects of the spherical accretion
rate by radiation pressure has to be expected.
Small changes in the temperature and density structure
in the inner cooling core region will directly have an effect on the accretion
rate. Therefore we have all the best prospects for building a
successful self-regulated AGN-feeding and cooling flow-heating model.

\section{Conclusions}

Several observational constraints have let us to the conclusion that
the mass deposition rates in galaxy cluster cooling cores are not
as high as previously predicted. The new X-ray spectroscopic observations
with a lack of spectral signatures for the coolest gas phases expected
for cooling flows and the lower mass deposition rates indicated at
other wavelength bands than X-rays are more consistent with mass
deposition rates reduced by one or two orders of magnitude below the previously
derived values. This can, however, only be achieved if the gas in the cooling
flow region is heated. The most
promising heating model is a self-regulated heating
model powered by the large energy output of the central AGN in
most cooling flows.
for cooling flows and the lower mass deposition rates indicated at
Most of the guidance and the support of the heating model proposed here
(based on concepts developed in Churazov et al. 2000, 2001)
is taken from the detailed
observations of a cooling core region in the halo of M87 and
to a smaller part from the observations in the Perseus cluster.
These observations show that the central AGN produces sufficient heat
for the energy balance of the cooling flow, that the most fundamental
and classical accretion process originally proposed by Bondi (1952)
provides an elegant way of devising a self-regulated model of 
AGN heating of the cooling flow, and that most of the further 
requirements that have to be met by a heating model to be consistent
with the observations can most probably be fulfilled.
Since these ideas are mostly developed to match the conditions in
M87, it is important to extent such detailed studies to most other
nearby cooling flow clusters.

In this new perspective the cooling cores of galaxy clusters become the
sites where most of the energy output of the central cluster AGN is
finally dissipated. Strong cooling flows should therefore be the
locations of AGN with the largest mass accretion rates. While in
the case of M87 with a possible current mass accretion rate of about
0.01 M$_{\odot}$ y$^{-1}$  the mass addition to the black hole
(with an estimated mass of about $3 \cdot 10^9$  M$_{\odot}$) 
is a smaller fraction of the total mass, the mass build-up may
become very important for the formation of massive black holes in
the most massive cooling flows, where mass accretion rates above
0.1 M$_{\odot}$ y$^{-1}$ become important over cosmological times.

{\it 
I like to thank the ROSAT team, the ESO key program team,
the NORAS team, and in particular Peter Schuecker, Luigi Guzzo,
Chris Collins, Kyoko Matsushita, Eugene Churazov, Yasushi Ikebe, 
Thomas Reiprich, and Yasuo Tanaka for the pleasant and fruitful
collaboration. I like to thank in particular Rashid Sunyaev for the
organization of this very magnificent conference !}

%INDEX%%%%%%%%%%%%%%%%%%%%%%%%%%%%%%%%%%%%%%%%%%%%%%%%%%%%%%%%%%%%%%%
% Please check with the editor of your book whether he plans to
% include a "mutual" subject index - if so, please code your entries
% in the standard syntax. For your own purposes you may print your
% "personal" index by using the following commands:
%
%\clearpage
%\addcontentsline{toc}{section}{Index}
%\flushbottom
%\printindex
%%%%%%%%%%%%%%%%%%%%%%%%%%%%%%%%%%%%%%%%%%%%%%%%%%%%%%%%%%%%%%%%%%%%%

\end{document}